\documentclass[11pt]{article}
\usepackage{amsmath,amsthm,amscd,array,amssymb}
\begin{document}
\date{}
\begin{center}
 {\Large {\bf BASIC PROPERTIES OF FEDOSOV SUPERMANIFOLDS}}\footnote{
 Contribution to Special Issue of Vestnik of Tomsk State
 Pedagogical University devoted to 70th Anniversary of Physical
 and Mathematical Department
 }
\end{center}

{\Large
\begin{center}
\medskip
{\sc B.~Geyer}$^{\ a)}$\footnote{E-mail: geyer@itp.uni-leipzig.de}
and
{\sc P.M.~Lavrov}$^{\ a), b)}$\footnote{E-mail: lavrov@tspu.edu.ru;
lavrov@itp.uni-leipzig.de}
\\
\vspace{1cm}

{\normalsize\it $^{a)}$ Center of Theoretical Studies, Leipzig University,\\
Augustusplatz 10/11, D-04109 Leipzig, Germany}
\vspace{.4cm}

{\normalsize\it $^{b)}$ Tomsk State Pedagogical University,
634041 Tomsk, Russia}
\end{center}
}
\vspace{.5cm}

\begin{quotation}
\setlength{\baselineskip}{12pt} \normalsize \noindent
Basic properties of even (odd) supermanifolds
endowed with a connection respecting a given symplectic
structure are studied.
Such supermanifolds can be considered as generalization
of Fedosov manifolds to the supersymmetric case.
\end{quotation}

\bigskip

\section{Introduction}
The formulation of fundamental physical theories, classical as well as quantum ones, by differential geometric methods nowadays is well established and has a great conceptual virtue. Probably, the most prominent example is the formulation of general relativity on Riemannian manifolds, i.e., the geometrization of gravitational force; no less important is the geometric formulation of gauge field theories of primary interactions on fiber bundles. Another essential route has been opened by the formulation of classical mechanics -- and also classical field theories -- on symplectic manifolds and their connection with geometric quantization. The properties of such kind of manifolds are widely studied.

Recently, some advanced methods of Lagrangian quantization involve more complicated manifolds namely the so-called Fedosov manifolds, i.e., symplectic manifolds equipped with a symmetric connection which respects the symplectic structure. Fedosov manifolds have been introduced for the first time in the framework of deformation quantization \cite{F}. The properties of Fedosov manifolds have been investigated in detail (see, e.g.,~Ref.~\cite{fm}). Especially, let us mention that for any Fedosov manifold the scalar curvature $K$ is trivial, $K=0$, and that the  specific relation $\omega_{ij,kl}=(1/3)R_{klij}$, in terms of normal coordinates, holds between the symplectic structure $\omega_{ij}$ and the curvature tensor $R_{klij}$.

The discovery of supersymmetry \cite{GL} enriched modern quantum field theory with the notion of supermanifolds being studied extensively by Berezin \cite{Ber}. Systematic considerations of supermanifolds and Riemannian supermanifolds were performed by DeWitt \cite{DeWitt}. At present, symplectic supermanifolds and the corresponding differential geometry are widely involved and studied in consideration of some problems of modern theoretical and mathematical physics \cite{bv,geom}.

However, the situation concerning Fedosov supermanifolds is quite different. Only flat even Fedosov supermanifolds have been used in the study of a
coordinate-free scheme of deformation quantization \cite{bt}, for an explicit realization of the extended antibrackets \cite{GrS} and for the formulation of the modified triplectic quantization in general coordinates, see, \cite{gl} and references cited therein. Here, on the basis previous results \cite{gl1,gl2}, we give an overview on the present status concerning the structure of arbitrary Fedosov supermanifolds, especially the properties of their curvature tensor and scalar curvature as well as the relations between the supersymplectic structure, the connection and the curvature in normal and general local coordinates.

The paper is organised as follows. In Sect.~2, we  give a brief review of the
definition of tensor fields on supermanifolds. In Sect.~3, we
consider affine connections on a supermanifold and their curvature tensors.
In Sect.~4, we present the notion of even (odd) Fedosov
supermanifolds and of even (odd) symplectic curvature tensors. In
Sect.~5, we study the introduction of the Ricci tensor and the property the
scalar curvature which is non-trivial for odd Fedosov supermanifolds.
In Sect.~6, we introduce normal coordinates on supermanifolds and affine extensions of the Christoffel symbols as well as tensor fields. In Sect.~7, we derive the relation existing between the first order affine extension of the Christoffel symbols and the curvature tensor for any Fedosov supermanifold both in normal coordinates and arbitrary local coordinates. In Sect.~8, we present a relation between the second order affine extension of the symplectic structure and the curvature tensor. In Sect.~9 we give a few concluding remarks.

We use the condensed notation suggested by DeWitt \cite{DeWitt}. Derivatives
with respect to the coordinates $x^i$ are understood as acting
from the left and for them the notation $\partial_i A={\partial
A}/{\partial x^i}$ is used. Right derivatives with respect to
$x^i$ are labelled by the subscript $"r"$ or the notation
$A_{,i}={\partial_r A}/{\partial x^i}$ is used. The Grassmann
parity of any quantity $A$ is denoted by $\epsilon (A)$.
\\

\date{}

\section{Tensor fields on supermanifolds}

To start with, we review explicitly some of the basic definitions and
simple relations of tensor analysis on supermanifolds which are
useful in order to avoid elementary pitfalls in the course of the
computations. Thereby, we adopt the conventions of DeWitt~\cite{DeWitt}.

Let the variables $x^i, \epsilon(x^i)=\epsilon_i$ be local
coordinates of a supermanifold $M, dim M=N,$ in the vicinity of a
point $P$. Let the sets $\{e_i:=\frac{\partial_r}{\partial x^i}\}$ and
$\{e^i:=dx^i\}$ be coordinate
bases in the tangent space $T_PM$ and the cotangent space
$T^*_PM$, respectively. If one goes over to another set ${\bar
x}^{i}={\bar x}^i(x)$ of local coordinates the basis vectors in
$T_PM$ and $T^*_PM$ transform as follows:
\begin{eqnarray}
\label{vec}
 {\bar e}_i=e_j \frac{\partial_r x^j}{\partial {\bar x}^i},
 \quad
{\bar e}^i=e^j \frac{\partial {\bar x}^i}{\partial x^j}.
\end{eqnarray}
For the transformation matrices the following relations hold:
\begin{eqnarray}
\label{unitJ}
 \frac{\partial_r {\bar x}^i}{\partial x^k}
 \frac{\partial_r x^k}{\partial {\bar x}^j}=\delta^i_j,
 \quad
 \frac{\partial x^k}{\partial {\bar x}^j}
 \frac{\partial {\bar x}^i}{\partial x^k}=\delta^i_j,
 \quad
 \frac{\partial_r x^i}{\partial {\bar x}^k}
 \frac{\partial_r {\bar x}^k}{\partial  x^j}=\delta^i_j,
 \quad
 \frac{\partial {\bar x}^k}{\partial  x^j}
 \frac{\partial  x^i}{\partial {\bar x}^k}=\delta^i_j.
\end{eqnarray}
Introduce the Cartesian product space $\Pi^n_m$
\begin{eqnarray}
\Pi^n_m=\overbrace{T^*_P\times \cdot\cdot\cdot\times T^*_P}^{n\;\; times}\times\underbrace{
T_P\times\cdot\cdot\cdot\times T_P}_{m\;\; times}.
\end{eqnarray}
Let ${\bf T}$ %($\epsilon({\bf T})=\epsilon(T)$)
be a mapping ${\bf T} : \Pi^n_m\rightarrow \Lambda$ that
sends every element
$(\omega^{i_1},..., \omega^{i_n}, X_{j_1},..., X_{j_m})\in \Pi^n_m$ into
supernumber
${\bf T}(\omega^{i_1},..., \omega^{i_n}, X_{j_1},..., X_{j_m})\in \Lambda$ where $\Lambda$ is the Grassmann algebra. This mapping is said to be a {\it tensor of rank (n,m) at P}
if for all $\omega,\sigma\in T^*_PM$, all $X,Y\in T_PM$ and all
$\alpha\in \Lambda$ it satisfies the multilinear laws
\begin{eqnarray}
\label{T}
\nonumber
{\bf T}(...\omega +\sigma...)&=&{\bf T}(...\omega...)+
{\bf T}(...\sigma...),\\
\nonumber
{\bf T}(...X +Y...)&=&{\bf T}(...X...)+{\bf T}(...Y...),\\
\nonumber
{\bf T}(...\omega\alpha, \sigma...)&=&{\bf T}(...\omega, \alpha\sigma...),\\
%\nonumber
{\bf T}(...\omega\alpha, X...)&=&{\bf T}(...\omega, \alpha X...),\\
\nonumber
{\bf T}(...X\alpha, Y...)&=&{\bf T}(...X, \alpha Y...),\\
\nonumber
{\bf T}(...X\alpha)&=&{\bf T}(...X) \alpha .
\end{eqnarray}
It is useful to work with
components of ${\bf T}$ relative to the bases $\{e^i\} and \{e_i\}$
\begin{eqnarray}
\label{cT}
%\nonumber
T^{i_1...i_n}_{\;\;\;\;\;\;\;\;\;j_1...j_m}=
{\bf T}(e^{i_1},...,e^{i_n}, e_{j_1},..., e_{j_m}),\quad
T_{j_1...j_m}^{\;\;\;\;\;\;\;\;\;i_1...i_n}=
{\bf T}(e_{j_1},..., e_{j_m},e^{i_1},...,e^{i_n}).
\end{eqnarray}
Then a tensor field of type $(n,m)$ with rank $n+m$ is defined as a
geometric object which, in each local coordinate system
$(x)=(x^1,...,x^N)$, is given by a set of functions with $n$ upper
and $m$ lower indices obeying definite transformation rules.
Here we omit the transformation rules for the components of any tensor
under a change of coordinates, $(x)\rightarrow ({\bar x})$,
referring to \cite{gl1}, and restrict ourselves to the case of the
second order tensor only. From (\ref{vec}), (\ref{T}) and (\ref{cT})
it follows
\begin{eqnarray}
\label{tr1}
{\bar T}^{ij}&=&
T^{mn}\frac{\partial {\bar x}^j}{\partial x^n}
\frac{\partial {\bar x}^i}{\partial x^m}
(-1)^{\epsilon_j(\epsilon_i+\epsilon_m)}, \quad
{\bar T}_{ij}=
T_{mn}\frac{\partial_r x^n}{\partial {\bar x}^j}
\frac{\partial_r x^m}{\partial {\bar x}^i}
(-1)^{\epsilon_j(\epsilon_i+\epsilon_m)},\\
\label{tr2}
{\bar T}^i_{\;j}&=& T^m_{\;\;\;n}
\frac{\partial_r x^n}{\partial {\bar x}^j}
\frac{\partial {\bar x}^i}{\partial x^m}
(-1)^{\epsilon_j(\epsilon_i+\epsilon_m)}, \quad
{\bar T}_i^{\;j}= T_m^{\;\;n}\frac{\partial
{\bar x}^j}{\partial x^n} \frac{\partial_r x^m}{\partial {\bar
x}^i} (-1)^{\epsilon_j(\epsilon_i+\epsilon_m)}.
\end{eqnarray}
Note that the unit matrix $\delta^i_j$ is connected with unit
tensor fields $\delta^i_{\;j}$ and $\delta^{\;\;i}_j$
%, transforming according to (\ref{form1}) and (\ref{form2}),
as follows
\begin{eqnarray}
\label{unit}
\delta^i_j=\delta^i_{\;j}=(-1)^{\epsilon_i}\;\delta^{\;\;i}_j =
(-1)^{\epsilon_j}\;\delta^{\;\;i}_j.
\end{eqnarray}
From a tensor field of type $(n,m)$ with rank $n+m$, where $n\neq
0, \;m\neq 0$, one can construct a tensor field of type
$(n-1,m-1)$ with rank $n+m-2$ by the contraction of an upper and a
lower index by the rules
\begin{eqnarray}
\label{tencontr} &&
 T^{i_1...i_{s-1}\;i \;i_{s+1}...i_n}
 _{\;\;\;\;\;\;\;\;\;\;\;\;\;\;\;\;\;\;\;\;\;\;\;\;\;\;\;\;\;\;\;
 j_1...j_{q-1}\;i\;j_{q+1}...j_m}\,
 (-1)^{\epsilon_i(\epsilon_{i_{s+1}}+\cdot\cdot\cdot +
 \epsilon_{i_n} + \epsilon_{j_1}+\cdot\cdot\cdot
 +\epsilon_{j_{q-1}}+1)},\\
\nonumber
\\
 \label{tencontr1} &&
 T_{ j_1...j_{q-1}\;i\;j_{q+1}...j_m}
 ^{\;\;\;\;\;\;\;\;\;\;\;\;\;\;\;\;\;\;\;\;\;\;\;\;\;\;\;\;\;\;\;
 i_1...i_{s-1}\;i \;i_{s+1}...i_n}\;
 (-1)^{\epsilon_i(\epsilon_{i_{s+1}}+\cdot\cdot\cdot +
 \epsilon_{i_n} + \epsilon_{j_1}+\cdot\cdot\cdot
 +\epsilon_{j_{q-1}})}.
\end{eqnarray}
In particular, for the tensor fields of type $(1,1)$ the
contraction leads to the supertraces,
\begin{eqnarray}
\label{sc1} T^i_{\;\;i}\;(-1)^{\epsilon_i}
 \quad {\rm and}\quad T^{\;\;i}_i.
\end{eqnarray}

>From two tensor fields $T^{i_1...i_n}$ and $P_{j_i...j_m}$ of types $(n.0)$ and $(0,m)$ one can construct new tensor fields of type $(n-1,m-1)$ using the multiplication procedure in the following way:
\begin{eqnarray}
\label{contr1}
 (-1)^{\epsilon(P)(\epsilon_{i_{1}}+\cdots +
 \epsilon_{i_{n-1}}+\epsilon_{k}) + \epsilon_{k}}\;
 T^{i_1...i_{n-1} k}\;P_{k j_1...j_{m-1}},\\
 \label{contr2}
 (-1)^{\epsilon(T)(\epsilon_{j_{1}}+\cdots +
 \epsilon_{j_{m-1}}+\epsilon_{k})}\;
 P_{j_1...j_{m-1} k}\;T^{k i_1...i_{n-1}}\;.
\end{eqnarray}
In particular, for the second rank tensor fields $T^{ij}$ and $P_{ij}$
begin{eqnarray}
\begin{eqnarray}
\label{contr2}
(-1)^{\epsilon(P)(\epsilon_i+\epsilon_k)+\epsilon_k}T^{ik}P_{kj}\;\;
{\rm or} \;\;(-1)^{\epsilon(T)(\epsilon_i+\epsilon_k)}P_{ik}T^{kj}.
\end{eqnarray}

Furthermore, taking into account (\ref{contr2}), the unique inverse of a (non-degenerate) second
rank tensor field of type (2,0) will be defined as follows:
\begin{eqnarray}
\label{invers1}
&&(-1)^{(\epsilon_i+\epsilon_k)\epsilon(T)+\epsilon_k}\;
T^{ik}\;(T^{-1})_{kj}
= \delta^i_{\;j},\quad
(-1)^{(\epsilon_j+\epsilon_k)\epsilon(T)}\;(T^{-1})_{jk}\;T^{ki}
= \delta_j^{\;\;i}\,,
\\
\nonumber \quad
 &&\epsilon(T^{-1}_{ij})=\epsilon(T^{ij})=
 \epsilon(T)+\epsilon_i+\epsilon_j\,,
\end{eqnarray}
and correspondingly for tensor fields of type (0,2).

%Of course, contractions between any number of tensor fields of
%arbitrary type are possible, but, here we do not need them.

Let us emphasize that the inclusion of the correct sign factors into
the definitions of contractions, (\ref{tencontr}) and
(\ref{tencontr1}), and of the inverse tensors, (\ref{invers1}), is essential. Namely,
%this is necessary
%concerning symmetry properties of tensor fields and their inverse
%ones on supermanifolds. For simplicity,
let us consider a second rank tensor field of type (2,0) obeying
the property of generalized (anti)symmetry,
\begin{eqnarray}
\label{sym}
 T_\pm^{ij}&=&\pm(-1)^{\epsilon_i\epsilon_j}T_\pm^{ji}. %,\\
\end{eqnarray}
Obviously, that property is in agreement with the transformation
law (\ref{tr1}),
\begin{eqnarray}
%\label{asym1}
\nonumber
&&{\bar T}_\pm^{ij}= T_\pm^{mn}\frac{\partial {\bar
x}^j}{\partial x^n} \frac{\partial {\bar x}^i}{\partial x^m}
(-1)^{\epsilon_j(\epsilon_i+\epsilon_m)}= \pm T_\pm^{nm}
\frac{\partial {\bar x}^i}{\partial x^m} \frac{\partial {\bar
x}^j}{\partial x^n} (-1)^{\epsilon_i\epsilon_n} =\pm
(-1)^{\epsilon_i\epsilon_j}{\bar T}_\pm^{ji}.
\end{eqnarray}
Thus, the notion of generalized (anti)symmetry of a tensor field
of type (2,0) is invariantly defined in any coordinate system.

Now, suppose that $T_\pm^{ij}$ is non-degenerate, thus allowing
for the introduction of the corresponding inverse tensor fields of
type (0,2) according to (\ref{invers1}). From
(\ref{sym}) one gets
\begin{eqnarray}
\label{inversym}
(T^{-1}_\pm)_{ij}=\pm(-1)^{\epsilon_i\epsilon_j+\epsilon(T)}(T^{-1}_\pm)_{ji}\,,
\end{eqnarray}
and, as it should be, also this generalized (anti)symmetry is
invariantly defined.
\\

\section{Affine connection on supermanifolds and curvature}

In analogy to the case of tensor analysis on manifolds, on a supermanifold
$M$ one introduces the covariant derivation (or affine
connection) as a mapping $\nabla$ (with components $\nabla_i,\,
\epsilon(\nabla_i)= \epsilon_i$) from the set of tensor fields on
$M$ to itself by the requirement that it should be a tensor
operation acting from the right and adding one more lower index
and, when it is possible locally to introduce Cartesian
coordinates on $M$, that it should reduce to the usual
(right--)differentiation.
For arbitrary supermanifolds the covariant derivative $\nabla $
(or connection $\Gamma$) is defined through the (right--)
differentiation and the separate contraction of upper and lower
indices with the connection components accompanig definite numerical factors which depend on the Grassmann parities of local coordinates. More explicitly, they are given as
local operations acting on scalar, vector and co-vector fields
%$T,\,T^i$ and $T_i$, respectively,
by the rules
\begin{eqnarray}
\label{scal} T\,\nabla_i&=&T_{,i}\,,
\\
\label{vector} T^i\,\nabla_j&=&T^i_{\;\;,j}+ T^k\Gamma^i_{\;\;kj}
(-1)^{\epsilon_k(\epsilon_i+1)}\,,
\\
T_i\,\nabla_j&=&T_{i,j}-T_k\Gamma^k_{\;\;ij}\,,
\end{eqnarray}
and on second-rank tensor fields of type $(2,0), (0,2)$ and
$(1,1)$ by the rules
\begin{eqnarray}
%\nonumber
{T}^{ij}\,{\nabla}_k&=& {T}^{ij}_{\;\;\;,k} +
{T}^{il}\,\Gamma^j_{\;\;lk}(-1)^{\epsilon_l(\epsilon_j+1)}+
{T}^{lj}\,\Gamma^i_{\;\;lk}
(-1)^{\epsilon_i\epsilon_j+\epsilon_l(\epsilon_i+\epsilon_j+1)}\,,\\
%\nonumber
{T}_{ij}\,{\nabla}_k&=& {T}_{ij,k} -
{T}_{il}\,\Gamma^l_{\;\;jk}-
{T}_{lj}\,\Gamma^l_{\;\;ik}
(-1)^{\epsilon_i\epsilon_j+\epsilon_l\epsilon_j}\,,\\
%\nonumber
{T}^i_{\;\;j}\,{\nabla}_k&=& { T}^i_{\;\;j,k} -
{T}^i_{\;\;l}\,\Gamma^l_{\;\;jk} +
{T}^l_{\;\;j}\,\Gamma^i_{\;\;lk}
(-1)^{\epsilon_i\epsilon_j+\epsilon_l(\epsilon_i+\epsilon_j+1)}\,.
%\nonumber
\end{eqnarray}
Similarly, the action of the covariant derivative on a tensor
field of any rank and type is given in terms of their tensor
components, their ordinary derivatives and the connection
components.

As usual, the affine connection components do not transform as mixed tensor
fields, instead they obtain an additional inhomogeneous term:
\begin{eqnarray}
{\bar\Gamma}^i_{\;\;jk}= (-1)^{\epsilon_n(\epsilon_m+\epsilon_j)}
 \frac{\partial_r \bar x^i}{\partial x^l}\Gamma^l_{\;\;mn}
 \frac{\partial_r  x^m}{\partial \bar x^j}
 \frac{\partial_r  x^n}{\partial \bar x^k}
 +
 \frac{\partial_r \bar x^i}{\partial x^m}
 \frac{\partial_r^2  x^m}{\partial \bar x^j \partial \bar x^k}\,.
\end{eqnarray}
In general, the connection components $\Gamma^i_{\;\;jk}$ do not
have the property of (generalized) symmetry w.r.t. the lower
indices. The deviation from this symmetry is the torsion,
\begin{eqnarray}
T^i_{\;\;jk} := \Gamma^i_{\;\;jk} -
(-1)^{\epsilon_j\epsilon_k}\Gamma^i_{\;\;kj}\,,
\end{eqnarray}
which transforms as a tensor field.
If the supermanifold $M$ is torsionless, i.e., $T^i_{\;\;jk}=0$,
then one says that a symmetric
connection is defined on $M$. Here, with the aim of studying
Fedosov supermanifolds, we consider only symmetric connections.

The Riemannian tensor field $R^i_{\;\;mjk}$, according to
Ref.~\cite{DeWitt}, is defined in a coordinate basis by the action
of the commutator of covariant derivatives, $[\nabla_i,\nabla_j]=
\nabla_i\nabla_j-(-1)^{\epsilon_i\epsilon_j}\nabla_j\nabla_i$, on
a vector field $T^i$ as follows:
\begin{eqnarray}
T^i[\nabla_j,\nabla_k]=-(-1)^{\epsilon_m(\epsilon_i+1)}
T^mR^i_{\;\;mjk}.
\end{eqnarray}
A straightforward calculation yields %the result
\begin{eqnarray}
\label{Ra} R^i_{\;\;mjk}=-\Gamma^i_{\;\;mj,k}+
\Gamma^i_{\;\;mk,j}(-1)^{\epsilon_j\epsilon_k}+
\Gamma^i_{\;\;jn}\Gamma^n_{\;\;mk}(-1)^{\epsilon_j\epsilon_m}-
\Gamma^i_{\;\;kn}\Gamma^n_{\;\;mj}
(-1)^{\epsilon_k(\epsilon_m+\epsilon_j)}.
\end{eqnarray}
The Riemannian tensor field possesses the following generalized
antisymmetry property,
\begin{eqnarray}
\label{Rsym}
R^i_{\;\;mjk}=-(-1)^{\epsilon_j\epsilon_k}R^i_{\;\;mkj}\,;
\end{eqnarray}
furthermore, it obeys the (super) Jacobi identity,
\begin{eqnarray}
\label{Rjac} (-1)^{\epsilon_m\epsilon_k}R^i_{\;\;mjk}
+(-1)^{\epsilon_j\epsilon_m}R^i_{\;\;jkm}
+(-1)^{\epsilon_k\epsilon_j}R^i_{\;\;kmj}\equiv 0\,
\end{eqnarray}
and the (super) Bianchi identity,
\begin{eqnarray}
\label{BI} (-1)^{\epsilon_i\epsilon_j}R^n_{\;\;mjk;i}
+(-1)^{\epsilon_i\epsilon_k}R^n_{\;\;mij;k}
+(-1)^{\epsilon_k\epsilon_j}R^n_{\;\;mki;j}\equiv 0\,,
\end{eqnarray}
with the notation $R^n_{\;\;mjk;i}:\,=R^n_{\;\;mjk}\nabla_i$.
\\

\section{Fedosov supermanifolds}

Suppose now we are given an even (odd) symplectic supermanifold,
$(M,\omega)$ with an even (odd) symplectic structure $\omega,\;
\epsilon(\omega) = 0$ (or $1$). Let $\nabla$ (or $\Gamma$) be a covariant derivative
(connection) on $M$ which preserves the 2-form $\omega$,
$\omega\nabla=0$.  In a coordinate basis this requirement reads
\begin{eqnarray}
\label{covomiv} \omega_{ij,k}-\omega_{im}\Gamma^m_{\;\;jk}+
\omega_{jm}\Gamma^m_{\;\;ik}(-1)^{\epsilon_i\epsilon_j}=0.
\end{eqnarray}
If, in addition, $\Gamma$ is symmetric then we have an even (odd)
symplectic connection (or symplectic covariant derivative) on $M$.
Now, a {\em Fedosov supermanifold} $(M,\omega,\Gamma)$ is
defined as a symplectic supermanifold with a given
symplectic connection.

Let us introduce the curvature tensor of a symplectic
connection with all indices lowered,
\begin{eqnarray}
\label{Rs}
R_{ijkl}=\omega_{in}R^n_{\;\;jkl},\quad
\epsilon(R_{ijkl})=\epsilon(\omega)+\epsilon_i+
\epsilon_j+\epsilon_k+\epsilon_l,
\end{eqnarray}
where $R^n_{\;\;jkl}$ is given by (\ref{Ra}). This leads to the
following representation,
\begin{eqnarray}
\label{Rse} R_{imjk}=-\omega_{in}\Gamma^n_{\;\;mj,k}+
\omega_{in}\Gamma^n_{\;\;mk,j}(-1)^{\epsilon_j\epsilon_k}+
\Gamma_{ijn}\Gamma^n_{\;\;mk}(-1)^{\epsilon_j\epsilon_m}-
\Gamma_{ikn}\Gamma^n_{\;\;mj}
(-1)^{\epsilon_k(\epsilon_m+\epsilon_j)}\,,
\end{eqnarray}
where we used the notation
\begin{eqnarray}
\label{G} \Gamma_{ijk}=\omega_{in}\Gamma^n_{\;\;jk},\quad
\epsilon(\Gamma_{ijk})=\epsilon(\omega)+
\epsilon_i+\epsilon_j+\epsilon_k\,.
\end{eqnarray}
Using this, the relation (\ref{covomiv}) reads
%\begin{eqnarray}
%\label{covom}
$\omega_{ij,k}=\Gamma_{ijk}-
\Gamma_{jik}(-1)^{\epsilon_i\epsilon_j}$.
%\end{eqnarray}
Furthermore, from Eq.~(\ref{Rse}) it is obvious that
\begin{eqnarray}
\label{Rans} R_{ijkl}=-(-1)^{\epsilon_k\epsilon_l}R_{ijlk},
\end{eqnarray}
and, using (\ref{Rs}) and (\ref{Rjac}), one deduces the (super)
Jacobi identity for $R_{ijkl}$,
\begin{eqnarray}
\label{Rjac1} (-1)^{\epsilon_j\epsilon_l}R_{ijkl}
+(-1)^{\epsilon_l\epsilon_k}R_{iljk}
+(-1)^{\epsilon_k\epsilon_j}R_{iklj}=0\,.
\end{eqnarray}

In addition, the curvature tensor $R_{ijkl}$ is (generalized)
symmetric w.r.t. the first two indices (see \cite{gl1}),
\begin{eqnarray}
\label{Ras}
R_{ijkl}=(-1)^{\epsilon_i\epsilon_j}R_{jilk}.
\end{eqnarray}

For any even (odd) symplectic connection there holds the identity
\begin{eqnarray}
\label{Rjac2}
(-1)^{\epsilon_i\epsilon_l}R_{ijkl}
+(-1)^{\epsilon_l\epsilon_k+\epsilon_l\epsilon_j}R_{lijk}
+(-1)^{\epsilon_k\epsilon_j+\epsilon_l\epsilon_j+\epsilon_i\epsilon_k}
R_{klij}+
(-1)^{\epsilon_i\epsilon_j+\epsilon_i\epsilon_k}R_{jkli}=0.
\end{eqnarray}
This is proved by using the Jacobi identity (\ref{Rjac1}) together
with a cyclic change of the indices.
In the identity (\ref{Rjac2}) the components of the symplectic
curvature tensor occur with cyclic permutations of all the
indices (on $R$). However, the pre-factors depending on the Grassmann
parities of indices are not obtained by cyclic permutation.

\section{Ricci and scalar curvature tensors}

Having  the curvature tensor, $R_{ijkl}$, and the tensor field
$\omega^{ij}$, which is inverse to $\omega_{ij}$,
\begin{eqnarray}
\label{A}
&&\omega^{ik}\;\omega_{kj}
(-1)^{\epsilon_k+\epsilon(\omega)(\epsilon_i+\epsilon_k)}=
\delta^i_j,\quad
(-1)^{\epsilon_i+\epsilon(\omega)(\epsilon_i+\epsilon_k)}
\omega_{ik}\;\omega^{kj}=\delta^j_i,\\
%\end{eqnarray}
%\begin{eqnarray}
\label{A11}
&&\omega^{ij}=-(-1)^{\epsilon_i\epsilon_j+\epsilon(\omega)}
\omega^{ji},\quad
\epsilon(\omega^{ij})= \epsilon(\omega)+\epsilon_i+\epsilon_j,
\end{eqnarray}
one can define the following three different tensor fields of type
$(0,2)$,
\begin{eqnarray}
\label{R1}
&&R_{ij}=\omega^{kn}R_{nkij}
 (-1)^{(\epsilon(\omega)+1)(\epsilon_k+\epsilon_n)}
 \qquad\;=\;R^k_{\;\;kij}\;(-1)^{\epsilon_k}
 \,,\\
\label{R2} &&K_{ij}=
\omega^{kn}R_{nikj}
 (-1)^{\epsilon_i\epsilon_k+(\epsilon(\omega)+1)(\epsilon_k+\epsilon_n)}
 \;=\;R^k_{\;\;ikj}\;(-1)^{\epsilon_k(\epsilon_i+1)}\,,\\
\label{R3}
&&Q_{ij}=\omega^{kn}R_{ijnk}
 (-1)^{(\epsilon_i+\epsilon_j)(\epsilon_k+\epsilon_n)+
 (\epsilon(\omega)+1)(\epsilon_k+\epsilon_n)}
 %\;=\;R^k_{\;\;ijk}(-1)^{\epsilon_k(\epsilon_i+\epsilon_j+1)}
 \,,\\
\nonumber &&\epsilon(R_{ij})=\epsilon(K_{ij})=\epsilon(Q_{ij})=
\epsilon_i + \epsilon_j\,.
\end{eqnarray}
From the definitions (\ref{R1}), (\ref{R3}) and the symmetry
properties of $R_{ijkl}$, it follows immediately  that for any
symplectic connection one has
$R_{ij}=-(-1)^{\epsilon_i\epsilon_j}R_{ji}$ and
$Q_{ij}=(-1)^{\epsilon_i\epsilon_j}Q_{ji}$. Moreover we obtain the
relations
%Therefore, on any even Fedosov supermanifold the tensor $R_{ij}$
%($Q_{ij}$) equals to zero (is non-trivial),  while on any odd
%Fedosov supermanifold this tensor is non-trivial (equals to zero).
%Indeed, using the symmetry properties of the tensor fields
%$\omega^{ij}$ and $R_{ijkl}$, one obtains the relations
\begin{eqnarray}
\label{R1p}
[1+(-1)^{\epsilon(\omega)}]R_{ij}=0,\quad
[1-(-1)^{\epsilon(\omega)}]Q_{ij}=0.
\end{eqnarray}
%In addition, for the tensor fields (\ref{R1}) -- (\ref{R3}) there
%exists a relation which follows from the identity (\ref{Rjac2}).
%Indeed, multiplying the equation (\ref{Rjac2}) by the factor $
%(-1)^{\epsilon_i\epsilon_l+(\epsilon(\omega)+1)(\epsilon_i+\epsilon_j)}
%$ and by the tensor $\omega^{ji}$, with allowance made for
%summation over indices $i,j$ and for the definitions (\ref{R1}) --
%(\ref{R3}), we obtain
>From (\ref{Rjac2}) and (\ref{R1}) -- (\ref{R3}) it follows the relations
\begin{eqnarray}
\label{Rl1}
&&R_{ij}+Q_{ij}+(-1)^{\epsilon_i\epsilon_j}K_{ji}+
(-1)^{\epsilon(\omega)}K_{ij}=0,\\
%\end{eqnarray}
%A second independent relation can be derived from (\ref{Rjac2}) by
%multiplying with the tensor $\omega^{ki}$ and the factor $
%(-1)^{\epsilon_i\epsilon_l+\epsilon_j\epsilon_k
%(\epsilon(\omega)+1)(\epsilon_i+\epsilon_j)} $; after subsequent
%summation over the indices $i,k$ this leads to the following
%result:
%\begin{eqnarray}
\label{Rl2}
&&[1+(-1)^{\epsilon(\omega)}]\left(K_{ij}-(-1)^{\epsilon_i\epsilon_j}K_{ji}\right)=0.
\end{eqnarray}
%From the relations (\ref{R1p}), (\ref{R3p}) and (\ref{Rl2}) one
%concludes:
Therefore  for any even symplectic connection we obtain
\begin{eqnarray}
\label{Rl3} K_{ij}=(-1)^{\epsilon_i\epsilon_j}K_{ji}, \quad
R_{ij}=0,\quad Q_{ij}=-2K_{ij},
\end{eqnarray}
while for any odd symplectic connection we have
\begin{eqnarray}
\label{Rl4} Q_{ij}=0,\quad
R_{ij}=K_{ij}-(-1)^{\epsilon_i\epsilon_j}K_{ji}.
\end{eqnarray}
The tensor field $K_{ij}$ should be considered as the
only independent second-rank tensor which can be constructed from
the symplectic curvature. We refer to $K_{ij}$ as the {\em Ricci tensor}
of an even (odd) Fedosov supermanifold. Notice that in the odd case
$K_{ij}$ has no special symmetry property.

Let us define the {\em scalar curvature} $K$ of a Fedosov supermanifold by the formula
\begin{eqnarray}
\label{Rsc} K=\omega^{ji}K_{ij}(-1)^{\epsilon_i+\epsilon_j}=
\omega^{ji}\omega^{kn}R_{nikj}
(-1)^{\epsilon_i+\epsilon_j+\epsilon_i\epsilon_k+
(\epsilon_k+\epsilon_n)(\epsilon(\omega)+1)}.
\end{eqnarray}
From the symmetry properties of $R_{ijkl}$ and $\omega^{ij}$, it
follows that on any Fedosov supermanifold one has
\begin{eqnarray}
\label{Rsc1}
[1+(-1)^{\epsilon(\omega)}]K=0.
\end{eqnarray}
Therefore, as is the case for ordinary Fedosov manifolds
\cite{fm}, for any even symplectic connection the scalar curvature
necessarily vanishes. But the situation becomes different for odd
Fedosov supermanifold where no restriction on the scalar curvature
occurs. Therefore, in contrast to both the usual Fedosov manifolds
and the even Fedosov supermanifolds, any odd Fedosov
supermanifolds can be characterized by the scalar curvature as an
additional geometrical structure \cite{gl1}. This basic property
of the scalar curvature can be used to formulate the following
statement \cite{r}: In both, the even and odd cases there exists
the relation $K^2=0$, and therefore any regular function of the
scalar curvature on any Fedosov supermanifolds belongs to class of
linear functions $\phi(x)=f[K(x)]=\alpha +\beta\, K(x)$ .

\section{Affine extensions of Christoffel symbols and tensors on
symplectic supermanifolds}

In Ref. \cite{fm} the virtues of using normal coordinates for
studying the properties of Fedosov manifolds was demonstrated.
Here, following Ref.~\cite{gl2}, we are going to extend this method
on Fedosov supermanifolds $(M, \omega , \Gamma)$ \cite{gl2}. Normal coordinates
$\{y^i\}$ within a point $p\in M$ can be introduced by using the  geodesic
equations as those local coordinates which satisfy the relations
($p$ corresponds to $y=0$)
\begin{eqnarray}
\label{Chsy}
 \Gamma^i_{\;jk}(y)\,y^k\,y^j = 0,
 \quad\epsilon(\Gamma_{ijk})=\epsilon(\omega)+\epsilon_i
+\epsilon_j+\epsilon_k.
\end{eqnarray}
It follows from (\ref{Chsy}) and the symmetry properties of $\Gamma_{ijk}$
w.r.t.~$(j\,k)$ that
\begin{eqnarray}
\label{Chsy0}
\Gamma_{ijk}(0)= 0.
\end{eqnarray}
In normal coordinates there exist additional relations at $p$ containing the partial derivatives of $\Gamma_{ijk}$. Namely, consider the Taylor expansion of $\Gamma_{ijk}(y)$ at $y=0$,
\begin{eqnarray}
\label{TCh}
\Gamma_{ijk}(y)=\sum_{n=1}^{\infty}\frac{1}{n!}
A_{ijkj_1...j_n}y^{j_n}\cdot\cdot\cdot y^{j_1},\quad
%\end{eqnarray}
{\rm where} \quad
%\begin{eqnarray}
%\label{Nt}
A_{ijkj_1\ldots j_n}=A_{ijkj_1\ldots j_n}(p)=\left.
\frac{\partial_r^n \Gamma_{ijk}}
{\partial y^{j_1}\ldots \partial y^{j_n}}\right|_{y=0}
\end{eqnarray}
is called an {\em affine extension} of $\Gamma_{ijk}$ of order $n=1,2,\ldots$ .
%Despite $\Gamma_{ijk}$ being not a tensor its affine extensions are true %tensors.
The symmetry properties of $A_{ijkj_1\ldots j_n}$ are evident from their definition (\ref{TCh}), namely, they are (generalized) symmetric w.r.t.~$(j\,k)$ as well as $(j_1\ldots j_n)$.
The set of all affine
extensions of $\Gamma_{ijk}$ uniquely defines a symmetric connection according to (\ref{TCh}) and satisfy an infinite sequence of identities \cite{gl2}.
In the lowest nontrivial order they have the form
\begin{eqnarray}
\label{r3}
A_{ijkl}+A_{ijlk}(-1)^{\epsilon_k\epsilon_l} +
A_{iklj}(-1)^{\epsilon_j(\epsilon_l+\epsilon_k)}=0.
\end{eqnarray}
%and
%\begin{eqnarray}
%\label{r6}
%\nonumber
%&&A_{ijklm}+
%A_{ijlkm}(-1)^{\epsilon_k\epsilon_l} +
%A_{ikljm}(-1)^{\epsilon_j(\epsilon_l+\epsilon_k)}\\
%&&+A_{ijmkl}(-1)^{\epsilon_m(\epsilon_l+\epsilon_k)}+
%A_{ilmjk}(-1)^{(\epsilon_j+\epsilon_k)(\epsilon_m+\epsilon_l)} +
%A_{ikmjl}(-1)^{\epsilon_j(\epsilon_m+\epsilon_k)+\epsilon_m\epsilon_l}=0%.
%\end{eqnarray}

Analogously, the affine extensions of an arbitrary tensor
$T=(T^{i_1...i_k}_{\;\;\;\;\;\;\;\;\;\;m_1...m_l})$ on $M$ are defined as  tensors on $M$ whose components at $p\in M$ in the local
coordinates $(x^1,\ldots,x^{2N})$ are given by the formula
\begin{eqnarray}
\label{AfT}
T^{i_1...i_k}_{\;\;\;\;\;\;\;\;m_1...m_l,j_1...j_n} \equiv
T^{i_1...i_k}_{\;\;\;\;\;\;\;\;m_1...m_l,j_1...j_n}(0)=\left.
\frac{\partial_r^n
T^{i_1...i_k}_{\;\;\;\;\;\;\;\;\;m_1...m_l}}
{\partial y^{j_1}...\partial y^{j_n}}\right|_{y=0}
\end{eqnarray}
where $(y^1,\ldots,y^{2N})$ are normal coordinates associated with
$(x^1,\ldots,x^{2N})$ at $p$. The first extension of any tensor
coincides with its covariant derivative because
$\Gamma^i_{\;\;jk}(0)=0$ in normal coordinates.

In the following, any relation containing affine extensions are to be
understood as holding in a neighborhood $U$ of an arbitrary point $p \in M$.
Let us also observe the convention that, if a relation holds for arbitrary
local coordinates, the arguments of the related quantities will be suppressed.

\section{First order affine extension of Christoffel symbols
and curvature tensor of Fedosov supermanifolds}

For a given Fedosov supermanifold
$(M,\omega, \Gamma)$, the symmetric connection $\Gamma$ respects the symplectic structure $\omega$ \cite{gl}:
\begin{eqnarray}
\label{sc}
\omega_{ij,k}=\Gamma_{ijk}-\Gamma_{jik}(-1)^{\epsilon_i\epsilon_j}.
\end{eqnarray}
Therefore, among the affine extensions of
$\omega_{ij}$ and $\Gamma_{ijk}$ there must exist some relations.
Introducing the affine extensions of $\omega_{ij}$
in the normal coordinates $(y^1,\ldots,y^{2N})$ at $p\in M$
according to,
\begin{eqnarray}
\label{Tew}
\omega_{ij}(y)
=\sum_{n=1}^{\infty}\frac{1}{n!} \Omega_{ij,j_1...j_n}\,
y^{j_n}\cdots y^{j_1},\quad
\Omega_{ij,j_1...j_n}=\omega_{ij,j_1...j_n}(0).
\end{eqnarray}
Using the symmetry properties of $\omega_{ij,j_1 \ldots j_n}(0)$ one
easily obtains the Taylor expansion for $\omega_{ij,k}$:
\begin{eqnarray}
\label{Tedw}
\omega_{ij,k}(y)
=\sum_{n=1}^{\infty}\frac{1}{n!}
\Omega_{ij,kj_1...j_n}\, y^{j_n}\cdots y^{j_1}.
\end{eqnarray}
Taking into account (\ref{sc}) and comparing
(\ref{TCh}) and (\ref{Tedw}) we obtain
\begin{eqnarray}
\label{wA}
\Omega_{ij,kj_1\ldots j_n}=A_{ijkj_1\ldots j_n}-
A_{jikj_1\ldots j_n}(-1)^{\epsilon_i\epsilon_j};
\end{eqnarray}
in particular,
\begin{eqnarray}
\label{wA1}
\Omega_{ij,kl}=A_{ijkl}-
A_{jikl}(-1)^{\epsilon_i\epsilon_j}.
\end{eqnarray}

Now, consider the curvature tensor $R_{ijkl}$
in the normal coordinates at $p\in M$. Then, due to $\Gamma_{ijk}(p)=0$, we obtain the following representation of the curvature tensor in terms of the affine extensions of the symplectic connection
\begin{eqnarray}
\label{Rnc}
R_{ijkl}(0)= -
A_{ijkl}+A_{ijlk}(-1)^{\epsilon_k\epsilon_l}.
\end{eqnarray}
Taking into account (\ref{r3}) %, (\ref{wA1})
and (\ref{Rnc})
a relation containing the curvature
tensor and the first affine extension of $\Gamma$ can be derived
Indeed, %from (\ref{Rnc}) and (\ref{r3})
the desired relation obtains as follows
\begin{eqnarray}
\label{RA1}
A_{ijkl}\equiv \Gamma_{ijk,l}(0) =
-\frac{1}{3}\left[R_{ijkl}(0)+R_{ikjl}(0)(-1)^{\epsilon_k\epsilon_j}\right],
\end{eqnarray}
where the antisymmetry (\ref{Rans}) of the curvature tensor
were used.

Notice, that relation (\ref{RA1}) was derived in normal coordinates.
It seems to be of general interest to find its analog relation
in terms arbitrary local coordinates $(x)$ because the Christoffel symbols
are not tensors while the r.h.s. of (\ref{RA1}) is a tensor.
Under that change of coordinates
$(x)\rightarrow (y)$ in some vicinity $U$ of $p$ the Christoffel symbols $\Gamma_{ijk}$ transform according to the rule
\begin{eqnarray}
\label{Chtr0}
%\nonumber
\Gamma_{ijk}(y)=
\left(\Gamma_{pqr}({x})
\frac{\partial_r {x}^r}{\partial y^k}
\frac{\partial_r {x}^q}{\partial y^j}
%\frac{\partial_r {x}^p}{\partial y^i}
(-1)^{\epsilon_k(\epsilon_j+\epsilon_q)}
+ \;\omega_{pq}({x})
\frac{\partial^2_r {x}^q}{\partial y^j\partial y^k}\right)
\frac{\partial_r {x}^p}{\partial y^i}
(-1)^{(\epsilon_k+\epsilon_j)(\epsilon_i+\epsilon_p)}.
\end{eqnarray}

In its turn the matrix of second derivatives
can be expressed in the form
\begin{eqnarray}
\label{Chtr1}
\frac{\partial^2_r {x}^q}{\partial y^j\partial y^k}=
\frac{\partial_r {x}^q}{\partial y^l}\Gamma^l_{\;\;jk}(y)-
\Gamma^q_{\;\;lm}({x})\frac{\partial_r {x}^m}{\partial y^k}
\frac{\partial_r {x}^l}{\partial y^j}
(-1)^{\epsilon_k(\epsilon_j+\epsilon_l)}.
\end{eqnarray}
In particular at $p\in M$ ($y=0$) we have the relation
\begin{eqnarray}
\label{Chtr2}
\left(\frac{\partial^2_r x^q}{\partial y^j\partial y^k}\right)_0
=
- \Gamma^q_{\;\;lm}(x_0)
\left(\frac{\partial_r x^m}{\partial y^k}\right)_0
\left(\frac{\partial_r x^l}{\partial y^j}\right)_0
(-1)^{\epsilon_k(\epsilon_j+\epsilon_l)}
\equiv - \Gamma^p_{\;\;jk}(x_0).
\end{eqnarray}

Differentiating (\ref{Chtr0}) with respect to $y$ we find
\begin{align}
\label{Chtr4}
\nonumber
\Gamma_{ijk,l}(y)
&=
\;\Gamma_{pqr;s}(x)
\frac{\partial_r x^s}{\partial y^l}
\frac{\partial_r x^r}{\partial y^k}
\frac{\partial_r x^q}{\partial y^j}
\frac{\partial_r x^p}{\partial y^i}
(-1)^{(\epsilon_j+\epsilon_k+\epsilon_l)(\epsilon_i+\epsilon_p)+
(\epsilon_k+\epsilon_l)(\epsilon_j+\epsilon_q)
+\epsilon_l(\epsilon_k+\epsilon_r)}\\
\nonumber
&~~+ \omega_{pq,r}(x)
\frac{\partial_r x^r}{\partial y^l}
\frac{\partial^2_r x^q}{\partial y^j\partial y^k}
\frac{\partial_r x^p}{\partial y^i}
(-1)^{(\epsilon_k+\epsilon_j)(\epsilon_i+\epsilon_p)+
\epsilon_l(\epsilon_i+\epsilon_j+\epsilon_k+\epsilon_q+\epsilon_p)}\\
\nonumber
&~~+ \omega_{pq}(x)
\frac{\partial^2_r x^q}{\partial y^j\partial y^k}
\frac{\partial^2_r x^p}{\partial y^i\partial y^l}
(-1)^{(\epsilon_k+\epsilon_j)(\epsilon_i+\epsilon_p)}+
\omega_{pq}(x)
\frac{\partial^3_r x^q}{\partial y^j\partial y^k\partial y^l}
\frac{\partial_r x^p}{\partial y^i}
(-1)^{(\epsilon_j+\epsilon_k+\epsilon_l)(\epsilon_i+\epsilon_p)},
\end{align}
where the covariant derivative (for arbitrary local coordinates) is defined by
\begin{eqnarray}
\label{Chc}
%\nonumber
\Gamma_{pqr;s}
=
\Gamma_{pqr,s}- \Gamma_{pqn}\Gamma^n_{\;\;\;rs}
- \Gamma_{pnr}\Gamma^n_{\;\;\;qs}
  (-1)^{\epsilon_r(\epsilon_n+\epsilon_q)}
- \Gamma_{nqr}\Gamma^n_{\;\;\;ps}
  (-1)^{(\epsilon_r+\epsilon_q)(\epsilon_n+\epsilon_p)}.
\end{eqnarray}
Restricting to the point $p \in M$ we get
\begin{align}
\Gamma_{ijk,l}(0)
=
\left(\Gamma_{ijk;l}(x_0)-\Gamma_{iln}(x_0)\Gamma^n_{\;\;\;jk}(x_0)
(-1)^{\epsilon_l(\epsilon_j+\epsilon_k)}\right)+
\omega_{iq}(x_0)
\left(\frac{\partial^3_r x^q}{\partial y^j\partial y^k\partial y^l}\right)_0.
%(-1)^{\epsilon_i\epsilon_q}.
\end{align}
Due to (\ref{Chtr4}) and the identity (\ref{r3}),
the matrix of third derivatives at $p$ obeys the following relation,
\begin{align}
\label{3w1}
\nonumber
\omega_{iq}(x_0)
\left(\frac{\partial^3_r x^q}{\partial y^j\partial y^k\partial y^l}\right)_0
&=
- \frac{1}{3}
 \Big[
  \left(\Gamma_{ijk;l} - \Gamma_{ijn}\Gamma^n_{\;\;\;kl} \right)
  (-1)^{\epsilon_j \epsilon_l}
  +
  \left(\Gamma_{ilj;k} - \Gamma_{iln}\Gamma^n_{\;\;\;jk} \right)
  (-1)^{\epsilon_l \epsilon_k}
    \\
&\qquad \;
  +
  \left(\Gamma_{ikl;j} - \Gamma_{ikn}\Gamma^n_{\;\;\;lj} \right)
  (-1)^{\epsilon_k \epsilon_j}
\Big](x_0) \;(-1)^{\epsilon_j \epsilon_l}.
\end{align}
With the help of (\ref{3w1}) we get the following transformation law for $\Gamma_{ijk;l}$
under change of coordinates at the point $p$
\begin{align}
\label{Chtr5}
\Gamma_{ijk,l}(0)=
 \Big[\Gamma_{ijk,l}(x_0)-\frac{1}{3}Z_{ijkl}(x_0)\Big],
\end{align}
with the abbreviation
\begin{eqnarray}
\label{Zx}
%\nonumber
Z_{ijkl}&=&
\Gamma_{ijk;l}+\Gamma_{ijl;k}(-1)^{\epsilon_k\epsilon_l}+
\Gamma_{ikl;j}(-1)^{(\epsilon_k+\epsilon_l)\epsilon_j}\\
\nonumber
&&
+2\Gamma_{iln}\Gamma^n_{\;\;\;jk}
(-1)^{(\epsilon_k+\epsilon_j)\epsilon_l}-
\Gamma_{ikn}\Gamma^n_{\;\;\;jl}(-1)^{\epsilon_j\epsilon_k}-
\Gamma_{ijn}\Gamma^n_{\;\;\;kl}
\end{eqnarray}
In straightforward manner one can check that the relations (\ref{Chtr4})
reproduce the correct transformation law for the curvature tensor $R_{ijkl}$.
Therefore, relation (\ref{RA1}) is to be generalized as
\begin{eqnarray}
\label{RAx}
\Gamma_{ijk;l}-\frac{1}{3}Z_{ijkl}=
-\frac{1}{3}[R_{ijkl}+R_{ikjl}(-1)^{\epsilon_j\epsilon_k}].
\end{eqnarray}
The last equation gets an identity when using the definition (\ref{Zx}) of
$Z_{ijkl}(x)$ and relation for $R_{ijkl}(x)$ on the r.h.s..

\section{Second and third order affine extension of symplectic structure
and curvature tensor on Fedosov supermanifolds}

Now, let us consider the relation between the second order affine
extension of symplectic structure and the symplectic curvature
tensor. It is easily found by taking into account (\ref{wA1}) and
(\ref{RA1}). Indeed, the Jacobi identity (\ref{Rjac1}), we obtain
\begin{align}
\label{wAR1} \nonumber \omega_{ij,kl}(0)=A_{ijkl}-
A_{jikl}(-1)^{\epsilon_i\epsilon_j} =\frac{1}{3}R_{klij}(0)
(-1)^{(\epsilon_i+\epsilon_j)(\epsilon_k+\epsilon_l)},
\end{align}
Again, since $p\in M$ is arbitrary, we finally obtain its
generalization for any local coordinates $x$:
\begin{eqnarray}
\label{wARx}
\omega_{ij,k;l}-\frac{1}{3}\left[Z_{ijkl}-Z_{jikl}
(-1)^{\epsilon_i\epsilon_j}\right]=
\frac{1}{3}(-1)^{(\epsilon_i+\epsilon_j)(\epsilon_k+\epsilon_l)}
R_{klij}.
\end{eqnarray}

Furthermore, using the second Bianchi identity \cite{gl1} one gets a
relation between the first derivative of the curvature
tensor and the affine connections,
\begin{eqnarray}
\label{RA2}
R_{ijkl,m}=-A_{ijklm} + A_{ijlkm}(-1)^{\epsilon_l\epsilon_k}.
\end{eqnarray}
as well as the third affine extension of the symplectic structure
\begin{align}
%\label{wA3R}
\nonumber
\omega_{ij,klm}=&-\frac{1}{6}[R_{ikjl,m}(-1)^{\epsilon_j\epsilon_k}+
R_{ikjm,l}(-1)^{\epsilon_j\epsilon_k+\epsilon_m\epsilon_l}+
R_{iljm,k}(-1)^{\epsilon_j\epsilon_l+\epsilon_k(\epsilon_l+\epsilon_m)}-
\\
&-R_{jkim,l}(-1)^{\epsilon_i(\epsilon_k+\epsilon_j)}-
R_{jkim,l}(-1)^{\epsilon_m\epsilon_l+\epsilon_i(\epsilon_j+\epsilon_k)}-
R_{jlim,k}(-1)^{\epsilon_k(\epsilon_m+\epsilon_l)
+\epsilon_i(\epsilon_j+\epsilon_l)}].
\nonumber
\end{align}
In local coordinates $(x)$ the following identity can be proven:
\begin{align}
%\label{IIRx}
%\nonumber
&R_{mjik;l}(-1)^{\epsilon_j(\epsilon_i+\epsilon_k)}
-R_{mijl;k}(-1)^{\epsilon_k(\epsilon_l+\epsilon_j)}
%\\&
+R_{mkjl;i}(-1)^{\epsilon_i(\epsilon_j+\epsilon_k+\epsilon_l)}
-R_{mlik;j}(-1)^{\epsilon_l(\epsilon_i+\epsilon_j+\epsilon_k)}=0.
\nonumber
\end{align}
For the derivation of these relations, see, Ref.~\cite{gl2}.

\section{Summary}

We have considered  some properties of tensor fields defined on
supermanifolds $M$. It was shown that only the generalized
(anti)symmetry of tensor fields has an invariant meaning, and that
differential geometry on supermanifolds should be constructed in
terms of such tensor fields.

Any supermanifold $M$ can be equipped with a symmetric connection
$\Gamma$ (covariant derivative $\nabla$). The Riemannian tensor
$R^i_{\;\;jkl}$ corresponding to this symmetric connection
$\Gamma$ satisfies both the (super) Jacobi identity and the
(super) Bianchi identity.

Any even (odd) symplectic supermanifold can be equipped with a
symmetric connection respecting the given symplectic structure.
Such a symmetric connection is called a symplectic connection. The
triplet $(M,\omega,\Gamma)$ is called an even (odd) Fedosov
supermanifold. The curvature tensor $R_{ijkl}$ of a symplectic
connection obeys the property of generalized symmetry with respect
to the first two indices, and the property of generalized
antisymmetry with respect to the last two indices. The tensor
$R_{ijkl}$ satisfies the Jacobi identity and the specific (for the
symplectic geometry) identity (see (\ref{Rjac2})) containing the
sum of components of this tensor with a cyclic permutation of all
the indices, which, however, does not  (!) contain cyclic permuted
factors depending on the Grassmann parities of the indices.

On any even (odd) Fedosov manifold, the Ricci tensor $K_{ij}$ can
be defined. In the even case, the Ricci tensor obeys the property
of generalized symmetry and gives a trivial result for the scalar
curvature. On the contrary, in the odd case the scalar curvature,
in general, is nontrivial.

Using normal coordinates on a supermanifold equipped with a
symmetric connection we have found relations among the first order
affine extensions of the Christoffel symbols and the curvature
tensor, the second order affine extension of symplectic structure
and the curvature tensor. In similar way  it is possible to find
relations containing higher order affine extensions of sypmlectic
structure, the Christoffel symbols and the curvature tensor. We
have established the form of the obtained relations in any local
coordinates (see (\ref{RAx}), (\ref{wARx})). It was shown that
$\Gamma_{ijk;l}(x)-1/3Z_{ijkl}(x)$ is a tensor field in terms of
which the relations obtained for general local coordinates can be
presented, cf. Eq.~(\ref{Chtr5}).
\medskip

\noindent
{\sc Acknowledgements:}~
P.L. thanks Leipzig University, Graduate College Quantum Field
Theory, for kind hospitality. The work was supported under grant
DFG 436 RUS 17/15/04.

%\newpage

\end{document}